\documentclass[prl,twocolumn,showpacs,amsmath,amssymb]{revtex4}

\usepackage{bm}
\usepackage{dcolumn}
\usepackage[dvipdfm]{graphicx}

\usepackage{color}

\def \braket<#1>{\langle{#1}\rangle}
\newcommand  {\Tr}{\mathop{\mathrm{Tr}}\nolimits}

\renewcommand{\_}[1]  {_{\rm #1}}
\renewcommand{\^}[1]  {^{\rm #1}}
\def \c...{\mbox{\small $\cdots$}}

\begin{document}

\preprint{preprint}

\title{Disordered Weak and Strong Topological Insulators}
\author{Koji Kobayashi$^1$}
\author{Tomi Ohtsuki$^1$}
\author{Ken-Ichiro Imura$^2$}
\affiliation{$^1$Department of Physics, Sophia University, Tokyo Chiyoda-ku 102-8554, Japan}
\affiliation{$^2$Department of Quantum Matter, AdSM, Hiroshima University, Higashi-Hiroshima 739-8530, Japan}

\date{\today}

\begin{abstract}
 A global phase diagram of disordered weak and strong topological insulators is established numerically.
 As expected, the location of the phase boundaries is renormalized by disorder, 
a feature recognized in the study of the so-called topological Anderson insulator.
 Here, we report unexpected quantization, {\it i.e.}, robustness against disorder 
of the conductance peaks on these phase boundaries.
 Another highlight of the work is 
on the emergence of two subregions in the weak topological insulator phase under disorder.
 According to the size dependence of the conductance,
the surface states are either robust or ``defeated'' 
in the two subregions.
 The nature of the two distinct types of behavior is further revealed by studying the Lyapunov exponents.
\end{abstract}
\pacs{
73.20.-r, 
73.20.Fz, 
71.23.-k, 
71.30.+h  
}
\maketitle

 Robustness against disorder is a defining property of the topological quantum phenomena.
 Depending on the degree of this robustness, 
three-dimensional (3D) $\mathbb{Z}_2$ topological insulators (TIs) 
\cite{FuKaneMele:3DTI,MooreBalents:3DTI,Roy:3DTI} 
are classified into strong and weak (STI and WTI).
 Bulk-surface correspondence implies that 
an STI exhibits a single helical Dirac cone that is protected, 
while a WTI manifests generally an even number (possibly null) 
of Dirac cones depending on the orientation of the surface \cite{FuKane:TopoNum}.

 Unusual robustness of Dirac electrons 
(especially in the case of a single Dirac cone) 
against disorder has been widely recognized 
in the study of graphene \cite{Nomura:Dirac, Bardarson:Dirac}.
 As a consequence of the absence of backward scattering \cite{Ando:BerryCNT}, 
the Dirac electrons do not localize.
 However, in the presence of valleys (even number of Dirac cones)
they do localize mediated by intervalley scatterings \cite{SuzuuraAndo:Gp}.
 Does this mean that an STI continues to be an STI in the presence of arbitrarily strong disorder, 
while a WTI simply collapses on the switching on of 
the short-ranged potential disorder that induces intervalley scattering?

\begin{figure}[tb]
  \begin{tabular}{c}
   \includegraphics[width=85mm]{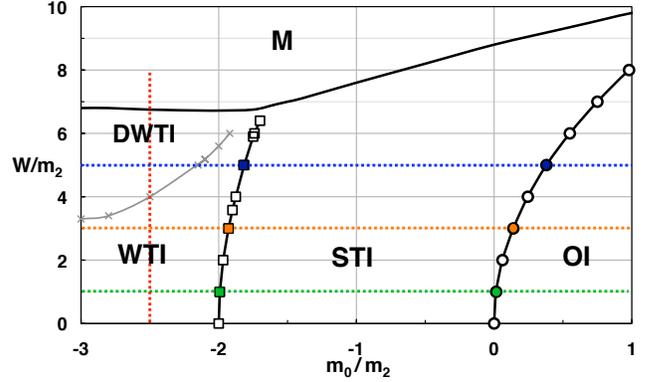}
  \end{tabular}
  \vspace{-2mm}
  \caption{ (Color online)
     The ``global phase diagram'' of the disordered $\mathbb{Z}_2$ topological insulator
    in the ($m_0/m_2$, $W/m_2$) plane determined by the behavior of two-terminal conductance.
     Solid lines on the phase boundaries are guides to the eyes.
     Dotted lines indicate the value of parameters relevant in Figs.~\ref{fig:G-m} and \ref{fig:G-W_m-25}.
     The metallic (M) phase lies in the intermediate range of disorder strength, typically $10\lesssim W/m_2 \lesssim 25$ in the parameter range of $m_0/m_2$ shown in this figure.
  }
  \label{fig:phaseD}
 \end{figure}

 Recent studies on the disordered WTI 
\cite{Mong:2PSWTI, Ringel:strongWTI} seem to suggest 
that the reality is much different.
 Our global phase diagram depicted in Fig.~\ref{fig:phaseD} 
finds its way also in this direction.
 This phase diagram is established by a combination of the study of 
the averaged two-terminal conductance 
and of the quasi-1D decay length 
in the transfer matrix approach.
 In the actual computation 
the 3D disordered $\mathbb{Z}_2$ topological insulator is modeled 
as an Wilson-Dirac-type tight-binding Hamiltonian with an effective ($\bm k$-dependent) mass term 
$m(\bm k) = m_0 + m_2 \sum_{\mu=x,y,z} (1-\cos k_\mu)$ \cite{Liu:3DTI}, 
implemented on a cubic lattice.
 The topological nature of the model is controlled by the ratio of two mass parameters
$m_0$ and $m_2$
such that
an STI phase with $\mathbb{Z}_2$ (one strong and three weak) indices \cite{FuKane:TopoNum} 
$(\nu_0, \nu_1 \nu_2 \nu_3)=(1, 000)$ 
appears when $-2 < m_0/m_2 < 0$,
while the regime of parameters $-4 < m_0/m_2 < -2$
falls on a WTI phase with $(\nu_0, \nu_1 \nu_2 \nu_3)=(0, 111)$ \cite{Imura:gappedTI}.

\begin{figure}[tb]
  \begin{tabular}{c}
   \includegraphics[width=80mm]{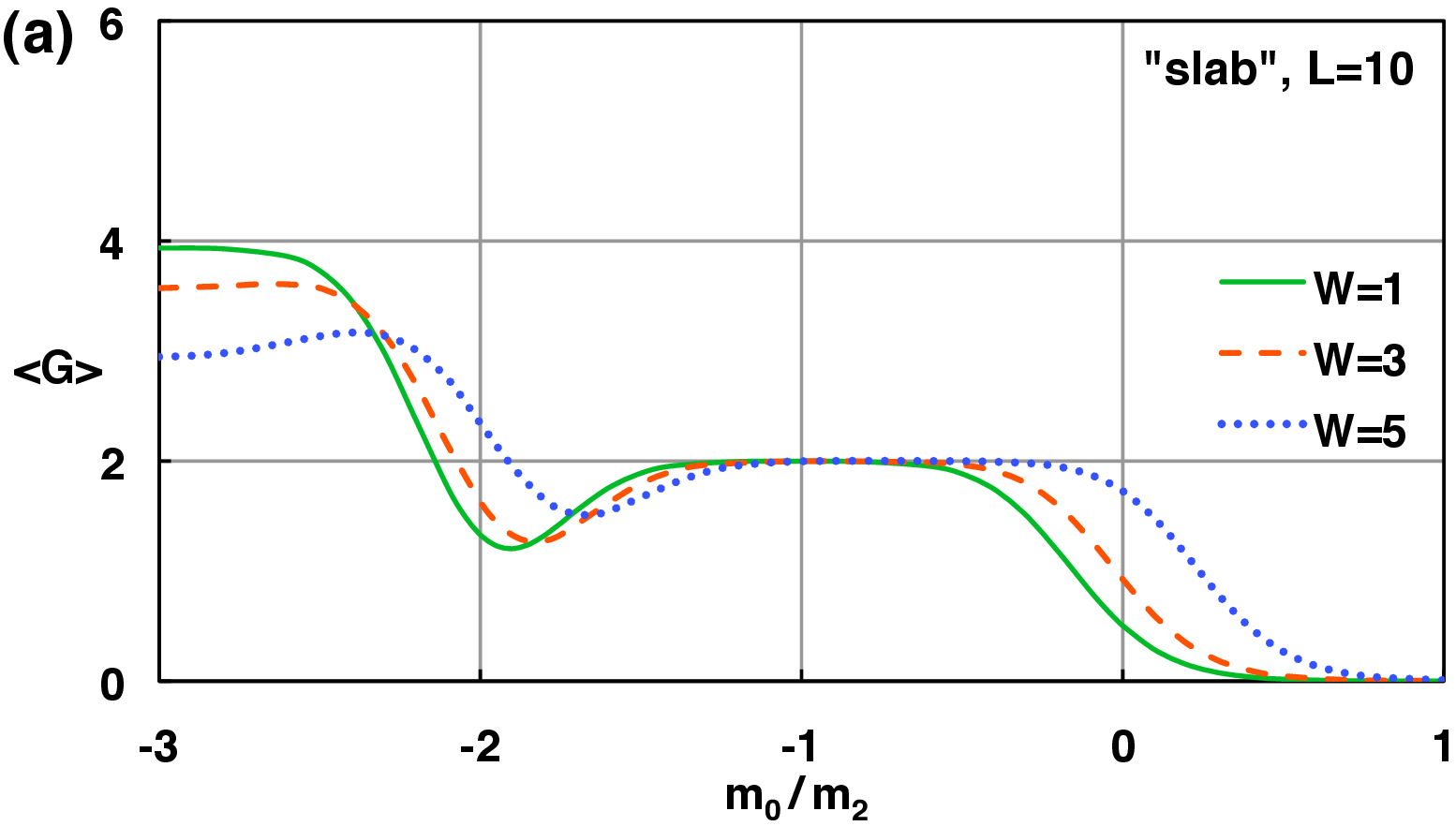}\\
   \includegraphics[width=80mm]{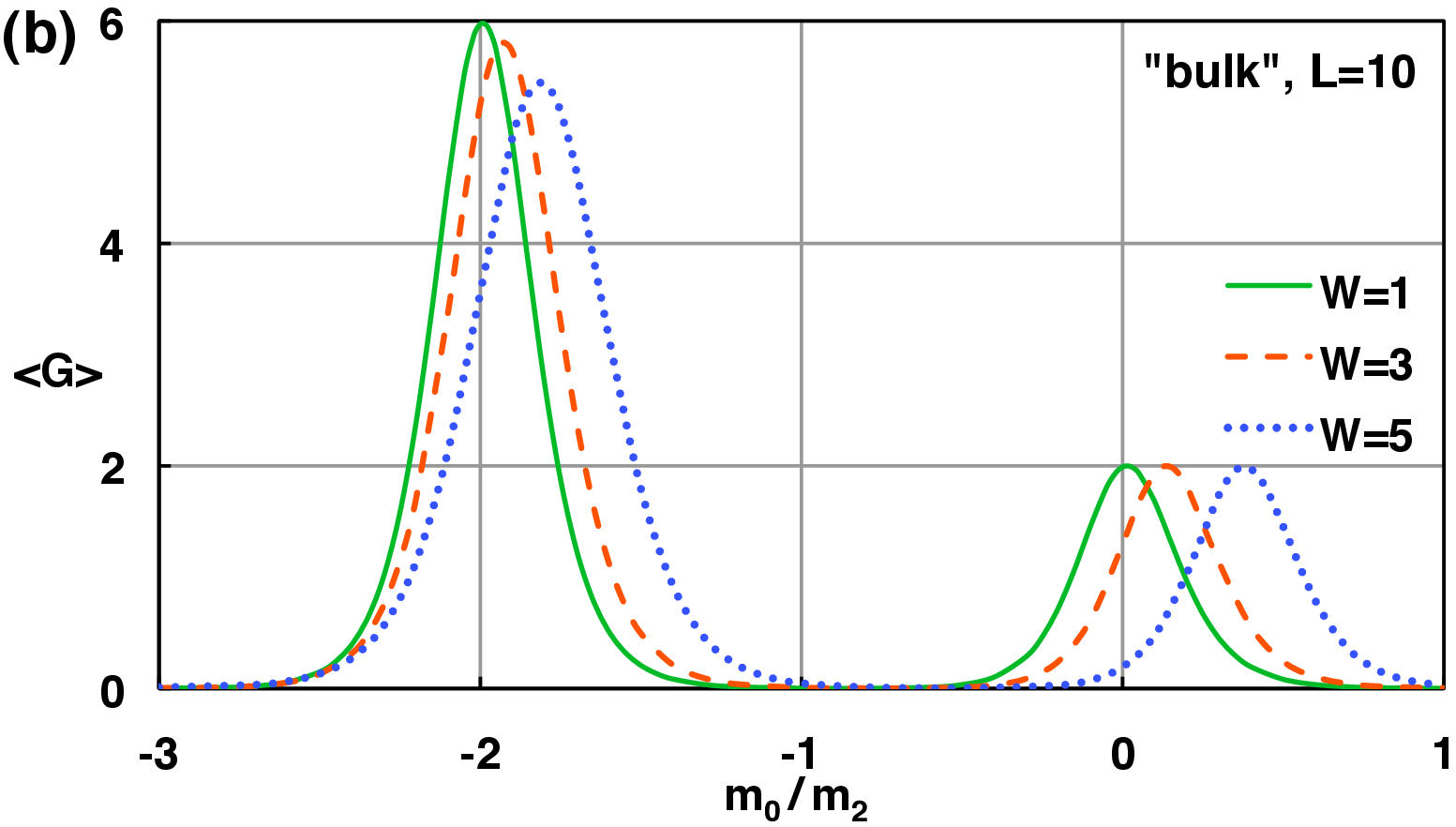}
  \end{tabular}
  \vspace{-2mm}
  \caption{ (Color online)
     Two-terminal conductance $\braket<G>$ (a) in the slab and (b) in the bulk geometries.
     $\braket<G>$ is plotted against $m_0/m_2$ for a system of size $L=10$ under different strength of disorder:
    $W=1$ (green solid line), $W=3$ (red dashed line), and $W=5$ (blue dotted line).
     In the bulk geometry PBCs are imposed in both the $x$ and $y$ directions,
    while in the slab geometry, an FBC is applied to the $x$ direction.
  }
  \label{fig:G-m}
 \end{figure}

 The obtained ``global'' phase diagram depicted in Fig.~\ref{fig:phaseD}
highlights the main results of the Letter.
 This phase diagram shows 
how disorder modifies the above topological classification in the clean limit
(naturally as a function of the strength of disorder $W$).
 To identify the nature of different phases and the location of the phase boundaries in the ($m_0/m_2$, $W/m_2$) plane, 
use of different geometries (i.e., bulk vs. slab) is shown to be crucial.
 While a plateau of the conductance in the slab geometry characterizes the nature of the corresponding phase [e.g., Fig.~\ref{fig:G-m}(a)], 
the phase boundaries are marked by a peak of the conductance in the bulk geometry [e.g., Fig.~\ref{fig:G-m}(b)].
 Under the breaking of translational invariance by disorder, 
standard techniques \cite{FuKane:TopoNum} for calculating the topological invariants fail.
 Yet, the above behaviors of the conductance clearly distinguish different topological phases,
providing us with sufficient information for establishing the phase diagram depicted in Fig.~\ref{fig:phaseD}.

 Both STI and WTI phases in the clean limit ($W=0$) survive in the presence of finite disorder, 
but collapse into a metallic phase (the M phase in Fig.~\ref{fig:phaseD}) at a finite strength of disorder.
 We have confirmed the robustness of these insulating phases 
by studying the system size dependence of the average conductance $\braket<G>$.
 The existence of surface helical Dirac cones and their nature are revealed 
by studying the generalized quasi-1D decay length $\xi_i$.
 A more specific comment on
the global structure of the phase diagram is 
that the STI exhibits a direct boundary with the ordinary insulator (OI) phase \cite{Shindo:3DZ2}, 
without being intervened on by an appearance of the metal phase.
 This is quite contrary to the case of the phase diagram for the 2D version of our model, 
in which the symplectic metal phase partitions the two insulating phases \cite{Yamakage:QSH, Prodan:QSH}.
 In the weakly disordered regime, 
a couple of marked features are to be mentioned.
 First the WTI phase has an internal structure; 
it is divided into WTI and defeated WTI (DWTI) regions, 
reflecting the change of the 
system size dependence of the conductance.
 This indicates that a WTI phase is in a sense indeed weak compared with an STI.
A detailed description of the DWTI region is given toward the end of the Letter.
 The second remark concerns 
the shape of the phase boundaries between different insulating phases.
 The positions of these phase boundaries are determined by the behavior of average conductance 
in the bulk and in the slab geometries (see Fig.~\ref{fig:G-m}).
 The location of the STI-OI boundary initially located at $m_0/m_2 =0$ in the clean limit
gradually moves toward 
the OI side while increasing disorder.
This feature
has been recognized in the study of the so-called ``topological Anderson insulator''
\cite{TAI1, TAI2, Guo:3DTAI}.
Here, we remark that a similar gradual shift of the phase boundary
also exists at the WTI-STI boundary.
On this WTI side, the STI turns out to be less expansive, invaded by the WTI.
This tendency is consistent with the result of SCBA calculation \cite{Nagoya}.


 As a concrete implementation of a 3D $\mathbb{Z}_2$ topological insulator on a lattice, 
we consider the following Wilson-Dirac-type tight-binding Hamiltonian \cite{Liu:3DTI},
   \begin{align} \label{eqn:H}
      H = & \sum_{\bf x} \sum_{\mu=x,y,z} \left[\frac{it}{2} c^{\dag}_{{\bf x}+{\bf e}_\mu} \alpha_{\mu} c_{\bf x}
                                         -\frac{m_2}{2}  c^{\dag}_{{\bf x}+{\bf e}_\mu} \beta c_{\bf x} + \rm{h.c.}\right]  \nonumber \\
            & + (m_0+3 m_2)\sum_{\bf x} c^{\dag}_{\bf x} \beta c_{\bf x}
            + \sum_{\bf x} v_{\bf x} c^{\dag}_{\bf x} 1_{4} c_{\bf x},
   \end{align}
where $c^{\dag}_{\bf x}$ and $c_{\bf x}$ are creation and annihilation operators on a site $\bf x$, 
${\bf e}_{\mu}$ is a unit vector, $\alpha_{\mu}$ and $\beta$ are gamma matrices in the Dirac representation 
   \begin{align} \label{eqn:gammaMat}
      \alpha_{\mu} = \begin{pmatrix}
                         0     & \sigma_\mu \\
                      \sigma_\mu &    0
                   \end{pmatrix}, \ 
      \beta = \begin{pmatrix}
                      1_{2} & 0 \\
                      0 & -1_{2}
                   \end{pmatrix}, 
   \end{align}
where $\sigma_{\mu}$ are Pauli matrices and $1_{2}$ is a $2\times 2$ identity matrix, 
$m_0, m_2$ and $t$ are mass and hopping parameters, 
and disorder potential $v_{\bf x}$ are uniformly distributed between $-W/2$ and $W/2$.
 For simplicity, we have assumed the Hamiltonian to be isotropic.
 We set, as in Ref.~\cite{RyuNomura:3DTI},
 $m_2 =1$, $t=2$ ($m_2$ is denoted as $r$ in Ref.~\cite{RyuNomura:3DTI}).
 This Hamiltonian belongs to the symmetry class AII \cite{Zirnbauer:UC}
(or, DIII for $W=0$).

 The transfer matrix \cite{MacKinnonKramer,EilmesRoemer:TMM} is given 
in terms of the wave function $\psi_n$ on a slice at $z=n$ as 
   \begin{align} \label{eqn:T} \nonumber
      \begin{pmatrix}
         \psi_{n+1} \\
         H_{+}\psi_{n}
      \end{pmatrix}
      = T_{n}
      \begin{pmatrix}
         \psi_{n} \\
         H_{+}\psi_{n-1}
      \end{pmatrix},
      T_{n} = 
      \begin{pmatrix}
         -H_{-}^{-1} H_{n}  &  -H_{-}^{-1}\\
         H_{+} & 0
      \end{pmatrix},
   \end{align}
where $H_{n} = \braket<n|H|n> - E$, $H_{-} = \braket<n|H|n+1>$ and $H_{+} = \braket<n+1|H|n>$.
 We set $E=0$ in this study, 
though similar results are obtained for $E=0.05$.

 To determine the phase boundaries between different insulating phases,
we calculate the (average) two-terminal conductance, 
using the Landauer formula \cite{Kobayashi:PGQSH}.
 The transport between the left and right terminals is described
in terms of the scattering matrix $\bm S$ defined as
   \begin{align}
      \begin{pmatrix}
         \psi\^{out}\_{L}  \\
         \psi\^{out}\_{R}
      \end{pmatrix}
      =\bm{S}
      \begin{pmatrix}
         \psi\^{in}\_{L}  \\
         \psi\^{in}\_{R}
      \end{pmatrix},
      \quad
      \bm{S}=
      \begin{pmatrix}
         \bm{r} & \bm{t}' \\
         \bm{t} & \bm{r}'
      \end{pmatrix},
   \end{align}
where $\psi\^{in(out)}\_{L(R)}$ denotes the incoming (outgoing) state on the left (right) terminal,
and $\bm{t}$ and $\bm{t}'$ ($\bm{r}$ and $\bm{r}'$) are transmission (reflection) matrices.
 The conductance $G$ in units of $e^2/h$ is given by $G = \Tr(\bm t^{\dag}\bm t)$.
 To find the explicit form of the $\bm S$-matrix, 
we employ a transfer matrix 
for a system of length $L$.
 The actual computation has been done 
in a cubic geometry of $L \times L \times L$ sites with electrodes attached to the $z$ direction.
 We assume that the electrodes consist of $L^2$ perfectly-conducting 1D wires 
as in network models \cite{Kobayashi:PGQSH}, 
so that 
details of the electrodes do not affect the conductance.

 To reinforce the validity of the phase diagram, 
we have also studied the Lyapunov exponents in the quasi-one-dimensional geometry of $L_x \times L_y \times \infty$.
 A Lyapunov exponent $\gamma_i$ is defined as
   \begin{align} \label{eqn:gamma}
    \gamma_i = \lim_{L_z\to\infty} \frac{\ln\lambda_i}{2M},
   \end{align}
where $\lambda_{i}$ is a positive eigenvalue of the matrix ${\cal T}^{\dag}{\cal T}$, 
and ${\cal T} = T_{M} \cdots T_{2} T_{1}$ is a product of the $M$ transfer matrices of dimension $8L_x L_y$ \cite{MacKinnonKramer}.
 Due to the current conservation, $\lambda_i$'s always come in reciprocal pairs
and due to the Kramers degeneracy, they are doubly degenerate.
 Keeping this in mind, 
we arrange the exponents in the decreasing order, 
$\gamma_{2L_x L_y} > \gamma_{2L_x L_y-1} > \cdots > \gamma_{2} > \gamma_{1} > 0 > -\gamma_{1} > \cdots > -\gamma_{2L_x L_y}$.
 The smallest positive Lyapunov exponent $\gamma_1$ is identified as the quasi-1D decay length $\xi$ by the correspondence, $\xi = 1/\gamma_1$. 
 Here, we extend this to higher Lyapunov exponents \cite{Slevin:higherLE,Kobayashi:LE2nd}:
   \begin{align} \label{eqn:Lambda}
    \xi_i = \frac{1}{\gamma_i}.
   \end{align}
 In the metallic phase, all the generalized decay length $\xi_i$'s 
increase monotonically as functions of the size of the cross section $L_x$ and $L_y$.
 In the OI phase, the decay lengths are insensitive to $L_x$ and $L_y$.
 In the topological insulating phases with slab geometry, 
higher decay lengths behave similarly to those of the OI phase, 
while the behavior of the largest few $\xi_i$'s is clearly distinguishable from those of the OI phase.
 They are significantly large, 
but in contrast to the metallic states, they show specific dependence on the system's thickness $L_x$.
 These behaviors reflect the nature of the 2D surface states.
 Indeed, the number of these atypical $\xi_i$'s corresponds to the number of Dirac cones on a surface, i.e., $1$ for STI and $2$ for WTI (DWTI) phases in Fig.~\ref{fig:phaseD}.


 Typical examples of the calculated conductance are shown in Fig.~\ref{fig:G-m}.
 In the upper panel (a) the conductance is calculated in the ``slab'' geometry, 
i.e., fixed boundary condition (FBC) in the $x$ direction and periodic boundary condition (PBC) in the $y$ direction (note $z$ direction is the direction of transport).
 The conductance shows a plateau behavior,
also quantized at $\braket<G>=2$ in the STI, 
and at $\braket<G>\simeq 4$ in the WTI phases.
 Note that near $E=0$, only the bands passing through the Dirac point contribute to the transport,
and the quantization of the conductance at $G=n$ implies that there are $n$ Dirac cones where the backscattering is strongly suppressed.
 In the lower panel (b), 
PBCs are imposed in both the $x$ and $y$ directions (``bulk'' geometry).
 The conductance is indeed very sensitive to the change of these boundary conditions.
 The prominent feature is a sharp peak 
at the phase boundary between different insulating phases, 
while inside the insulating phases, irrespective of their topological non-triviality (either in the STI, WTI, or OI phases), 
the conductance tends to vanish in the thermodynamic limit.
 The positions of the conductance peaks show little dependence on $L$ 
and have been used to determine the phase boundaries of Fig.~\ref{fig:phaseD}.

 In contrast to the expected quantized plateau due to the surface Dirac cones, 
the quantization of the peak height of conductance in the bulk geometry is unexpected [Fig.~\ref{fig:G-m}(b)].
 At the phase boundaries between different insulating phases, the bulk energy gap closes and in the model studied 3D Dirac cones emerge in the spectrum.
 The number of such Dirac cones is $1$ at the STI-OI boundary, 
while it is $3$ at the WTI-STI boundary.
 By studying the size dependence of $G$, 
we have verified that the peak height of $G$ approaches indeed a quantized value, 
i.e., $G=2$ at the STI-OI boundary, and $G=6$ at the WTI-STI boundary.
 Note that the quantized values are twice the number of corresponding Dirac cones.
 The factor of $2$ comes from two time-reversal pairs of conducting channels for each Dirac cone.

 The above situation of an even number of channels implies fluctuating nonquantized conductance due to disorder.
 As can be seen in Fig.~\ref{fig:G-m}(b), however, 
the height of the conductance peak is insensitive to the strength of disorder.
 We have also verified that the fluctuation of $G$ is strongly suppressed as approaching the peak.
 This is in contrast to the 2D TI transitions, where a finite universal fluctuation is observed \cite{Kobayashi:PGQSH}.
 This may be understood by noting that the disorder becomes irrelevant near the 3D TI transitions \cite{ShindouMurakami,GoswamiChakravarty}. (See also Supplemental Material \cite{supplemental}.)

 \begin{figure}[tb]
  \begin{tabular}{c}
   \includegraphics[width=80mm]{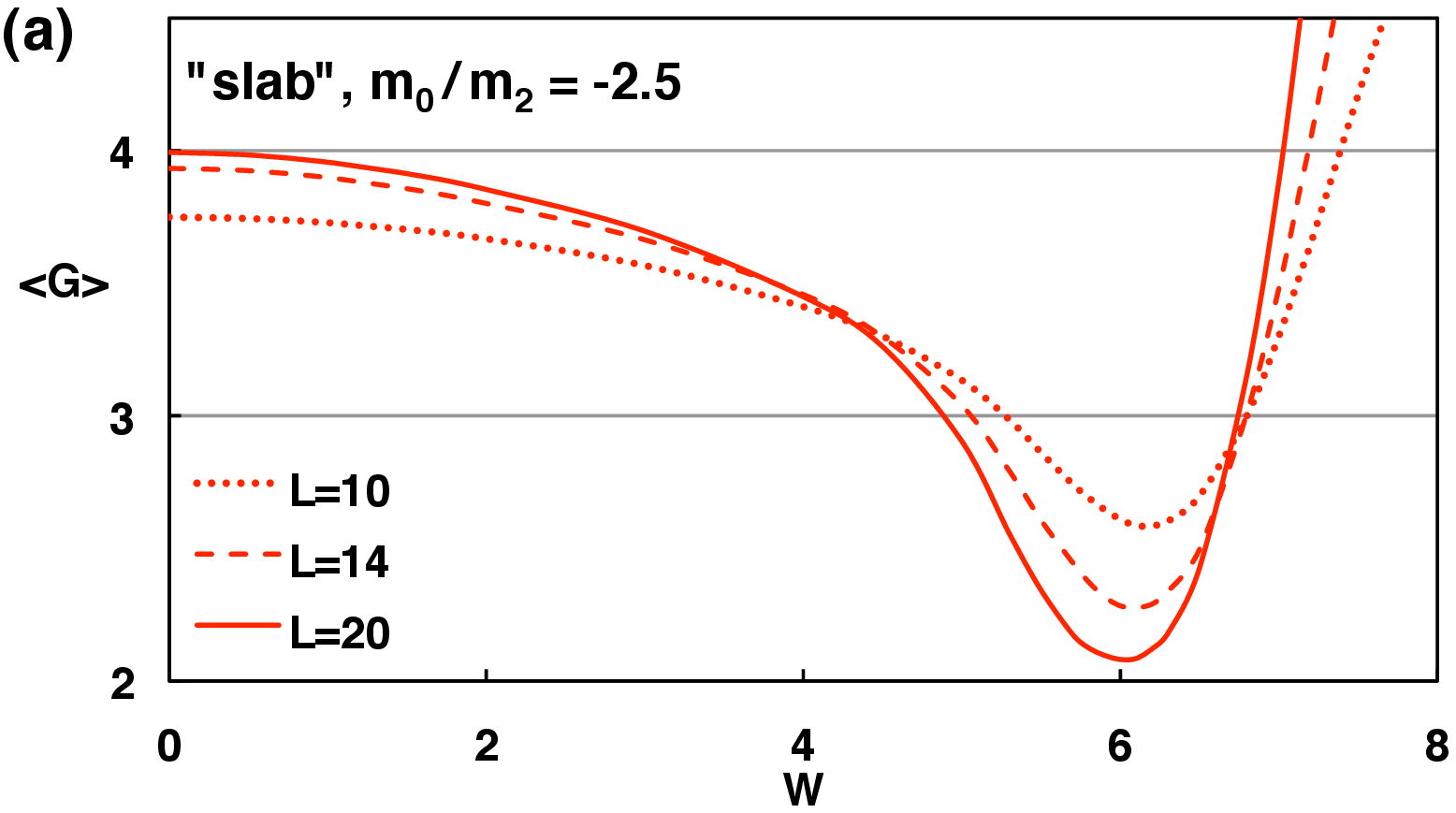}\\
   \includegraphics[width=80mm]{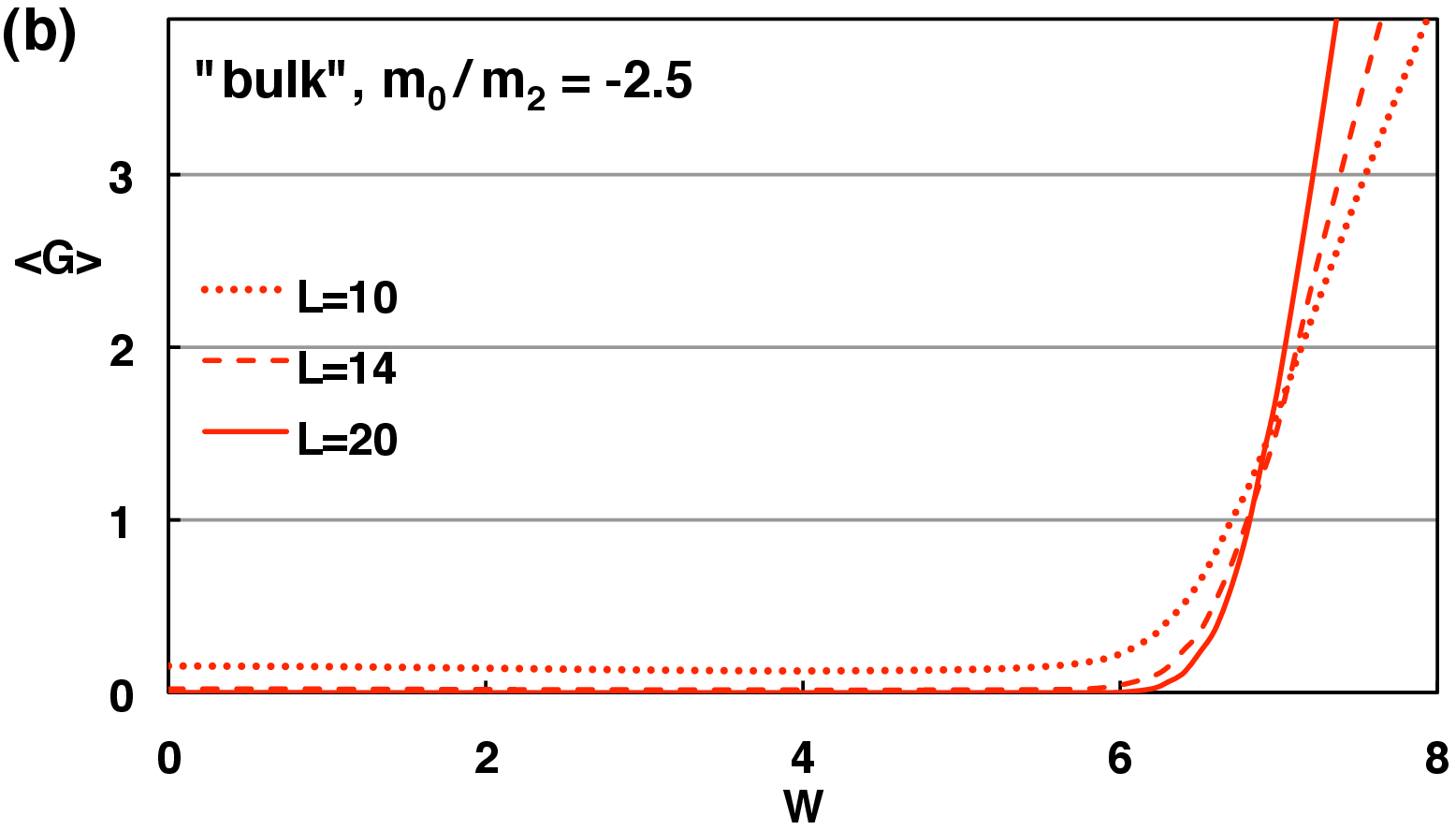}
  \end{tabular}
  \vspace{-2mm}
  \caption{ (Color online)
     Two-terminal conductance as a function of disorder with $m_0/m_2=-2.5$ for different system sizes: 
     $L=10$ (dotted line), $L=14$ (dashed line), and $L=20$ (solid line).
     The upper panel (a) is for the slab geometry.
     The WTI surface states are stable for a certain value of disorder (about $W\lesssim 4$, which is actually not small), and defeated by disorder around $W=6$.
     For $W\gtrsim 7$, the system enters the metallic phase.
     The lower panel (b) shows the conductance in bulk geometry.
     There are no peak, which should appear at the topological phase transition point around $W \simeq 4$.
     The statistical error is less than $0.02$.
  }
  \label{fig:G-W_m-25}
 \end{figure}

 \begin{figure}[tb]
  \begin{tabular}{cc}
   \includegraphics[width=80mm]{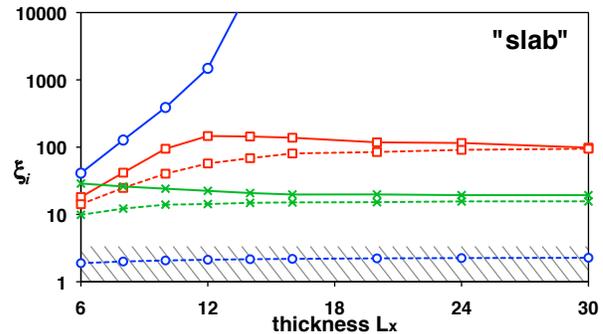}
  \end{tabular}
  \vspace{-2mm}
  \caption{ (Color online)
    Decay lengths $\xi_1$ (solid lines) and $\xi_2$ (dashed lines) 
   in the STI $(m_0/m_2,W/m_2)=(-1,1)$ (blue $\circ$), the WTI $(m_0/m_2,W/m_2)=(-2.5,3.5)$ (red {\tiny$\square$}), 
   and DWTI $(m_0/m_2,W/m_2)=(-2.5,6)$ (green $\times$) regions in the slab geometry with system width $L_y=6$.
    The higher decay lengths $\xi_{i\ge 3}$ appear in the shaded region, 
   which indicates the typical range of decay lengths in the OI phase.
    The error bars are less than $5\%$.
  }
  \label{fig:xi}
 \end{figure}

 Let us further quantify the transport characteristics
of the disordered STI and WTI phases (in Fig.~\ref{fig:phaseD})
by the analyses of the conductance and of the decay length.
 In the STI phase, 
the conductance approaches the quantized value $G=2$ irrespective of the disorder strength, 
and forms a conductance plateau [see Fig.~\ref{fig:G-m}(a)].
 In this plateau region, the largest decay length $\xi_1$ increases rapidly (exponentially) with $L_x$ (see, e.g., Fig.~\ref{fig:xi}), 
while the remaining $\xi_{i\ge 2}$ are
on the same order of their counterparts in the OI phase.
 This implies that there remains a single Dirac cone on each surface and 
in the limit $L_x \to \infty$, it becomes perfectly conducting, realizing effectively a system of an odd number of channels (see Refs.~\cite{Takane:SpOdd,Kobayashi:LT25}).

 On the other hand, in the WTI phase, the size dependence of the conductance 
implies that the WTI phase is divided into two qualitatively different regions; 
the $L$ dependence of the conductance shows a contrasting behavior
in the two regions.
 For weak disorder ($W\lesssim 4$ for $m_0/m_2=-2.5$), 
the conductance increases and asymptotically approaches a finite value $G\simeq 4$ as $L$ increases 
[see Fig.~\ref{fig:G-W_m-25}(a)].
 In this weakly disordered regime, 
the decay lengths $\xi_1$ and $\xi_2$ tend to increase as $\xi_1$ in the STI phase, 
before being saturated as $L_x$ increases (see Fig.~\ref{fig:xi}) at a value approximately $50$ times larger than $\xi_3$.
 The saturation of $\xi$ is caused by a small but finite intervalley scattering amplitude, which was absent in the STI phase.
 Combining these two observations, 
one can convince oneself that 
the surface states of a WTI indeed remain stable 
in this regime of a finite disorder strength.
 On the contrary, 
once the disorder exceeds a certain value ($W\simeq 4$), 
the conductance decays with increasing $L$ (Fig.~\ref{fig:G-W_m-25}) 
and $\xi_1$ and $\xi_2$ do not show an exponential rise.
 Such a behavior is indeed indistinguishable from the case of OI.
 What is then the nature of these two regimes, 
WTI and DWTI in Fig.~\ref{fig:phaseD}?
 Are they distinct phases, 
and is there a transition between the two?
 Our result suggests the following.
 The absence of a conductance peak between the WTI and DWTI regions [see Fig.~\ref{fig:G-W_m-25}(b)]
implies that they are topologically identical.
 Actually, $\xi_1$ and $\xi_2$, which correspond to the surface states, are still large and distinguishable from the bulk states (although they are smaller than those in the STI and WTI phases).
 In the DWTI region, 
surface conducting states are simply ``defeated'' by disorder.
 In this sense we name this region the defeated WTI or DWTI region.
 The fate of the WTI and DWTI regions in the thermodynamic limit is highly nontrivial.
 We leave a more substantial presentation on this point to a forthcoming publication.

 In this Letter, we have numerically investigated the transport property of 3D weak and strong topological insulators (WTI and STI).
 We have employed two approaches, 
i.e., two-terminal conductance and Lyapunov exponents,
to identify different phases and expanded the previously studied phase diagram, 
in particular, on the WTI side.
 We found that the WTI phase has an internal structure with an additional DWTI region.
 Although the configuration of different disordered topological phases 
and the absolute location of their phase boundaries are model (and in experimental terms, material) 
dependent,
the general tendency studied in this Letter about how these phase boundaries are renormalized by disorder is naturally presumed to be model independent.
 Last but not the least,
 the following two specific features
uncovered in this Letter,
(i) quantization of the peak conductance
characterizing these phase boundaries
to a universal value $2e^2/h$
(multiplied by the number of simultaneous gap closing),
(ii) the behavior of generalized decay length $\xi_i$
reflecting the number of surface Dirac cones,
are considered to be generic, and 
applicable to any alternative models of the WTI and STI.


\begin{acknowledgments}
 This work was supported by KAKENHI No.~23$\cdot$3743, No.~23540376, and No.~23103511.
 The authors thank Y. Takane, K. Nomura, K. Slevin, A. Yamakage, and R. Shindou for useful discussions.
\end{acknowledgments}

\end{document}